\documentclass{JHEP3}
\usepackage{verbatim}
\usepackage{epsfig}
\usepackage{bm}
\usepackage{amsthm,amsmath,amssymb,amsfonts}
\usepackage{cite}
\usepackage[english]{babel}

\newcommand{\beq}{\begin{equation}}
\newcommand{\eeq}{\end{equation}}
\newcommand{\bea}{\begin{eqnarray}}
\newcommand{\eea}{\end{eqnarray}}

\def\tv{{\widetilde{V}}}

\def\cA{{\cal A}}
\def\cE{{\cal E}}
\def\Pbr{{P\!\!\!\slash}}

\title{Parity Breaking Transport in Lifshitz Hydrodynamics}

\author{Carlos Hoyos$^1$, Adiel Meyer$^2$ and Yaron Oz$^2$\\
$^1$ Department of Physics, Universidad de Oviedo,
Avda. Calvo Sotelo 18, 33007, Oviedo, Spain\\
$^2$ Raymond and Beverly Sackler School of Physics and Astronomy, Tel Aviv University, Tel Aviv 69978, Israel\\
}

\abstract{
We derive the constitutive relations of first order charged hydrodynamics for theories with Lifshitz scaling and broken parity in $2+1$ and $3+1$ spacetime
dimensions.  In addition to the anomalous (in $3+1$) or Hall (in $2+1$) transport of relativistic hydrodynamics, there is an additional
non-dissipative transport allowed by the absence of boost invariance. We analyze the non-relativistic limit and use a phenomenological model of a strange metal to argue that these effects can be measured in principle 
by using electromagnetic fields with non-zero gradients.
}

\preprint{TAUP-2994/15, FPA-15/10}

\begin{document}


\section{Introduction}

Experiments suggest that quantum criticality is underlying many of the exotic transport properties of `strange metals' like high $T_c$ superconductors, heavy fermion, organic superconductors, etc. In most cases very little is known about the quantum critical theory or theories that are responsible for the observed behavior, including something as basic as their symmetries. Any clue in this direction would  then be extremely valuable. 

One of the simplest models of quantum critical behavior is the $z=2$ Lifshitz theory introduced in \cite{Ardonne:2003wa}. Theories with Lifshitz symmetry and dynamical exponent $z$ exhibit a minimal set of symmetries including time and space translations, space rotations and the anisotropic scaling $t\to \lambda^z t$, $x^i\to \lambda x^i$, which may be anomalous for some observables. Starting with \cite{Kachru:2008yh}, strongly coupled theories with Lifshitz symmetry have been extensively studied using gravity duals with the goal of identifying universal properties that would help to understand better the nature of the quantum critical point. 

However, Lifshitz symmetry can in principle be extended to larger groups that include Galilean boosts and for $z=2$ also non-relativistic conformal transformations \footnote{There are other possible extensions for discrete values of $z<1$ e.g.~\cite{Bagchi:2009my}.}. For $z=1$ it can be extended to the full relativistic conformal symmetry. One of the motivations for the present work is to identify effects that distinguish between systems with and without boost invariance. A first step in this direction is to develop the hydrodynamic description of a theory with Lifshitz symmetry at finite temperature. This is well motivated as an effective description of electrons in the case of strange metals \cite{Sachdev:2011}. The interaction rate between the electrons is much larger than the interaction rate with the ions in the lattice or impurities, so the electrons have time to reach local thermal equilibrium between scattering events with the other elements of the system that produce momentum loss. Note that this is not true in ordinary metals, that can be described as a weakly coupled gas of quasiparticles. 

The description of Lifshitz hydrodynamics was initiated in \cite{Hoyos:2013eza} for neutral fluids and extended for charged fluids and superfluids in \cite{Chapman:2014hja}. In the present work we extend the charged hydrodynamics to include parity breaking effects. This is partly motivated by the fact that in many cases the strange metal behavior occurs in a magnetic material and/or in the presence of magnetic fields. More generally, parity breaking introduces many  effects in hydrodynamics such as Hall transport in $2+1$ dimensions or anomalous transport in $3+1$ dimensions. An interesting question is if new effects are possible when there is no boost invariance. We find that this is indeed the case in $3+1$ dimensions, where there can be currents in the direction of magnetic fields or vorticity. Usually these kind of transports are associated to the presence of chiral anomalies and have been studied in multiple systems, including e.g.~astrophysics \cite{Vilenkin:1978is,Vilenkin:1980fu} quark-gluon plasma \cite{Fukushima:2008xe}, condensed matter \cite{Alekseev:1998,Volovik:1999wx}, cosmology \cite{Giovannini:1997eg} and relativistic hydrodynamics \cite{Son:2009tf,Neiman:2010zi,Sadofyev:2010pr,Landsteiner:2011cp},\footnote{In this context, anomalous transport coefficients were first computed in holographic models \cite{Newman:2005hd,Erdmenger:2008rm,Banerjee:2008th,Torabian:2009qk}.} where they are commonly known as Chiral Magnetic and Chiral Vortical Effects. The novelty is that in Lifshitz theories it is allowed even if the currents are non-anomalous. In $2+1$ dimensions there are additional terms that affect the Hall conductivity and Hall thermal transport.

The transport described above is in the context of non-invariance under Lorentz boosts. In addition, we take a non-relativistic limit to derive the Lifshitz hydrodynamics in the context of non-invariance under Galilean boosts. We use it as a phenomenological description of the strange metal as a Drude model with a strongly coupled electron fluid, in the spirit of \cite{Mendoza:2013} for graphene. We take into account the effects of scattering with the lattice or impurities by  adding a drag term to the hydrodynamic equations. Interestingly, we find that in $3+1$ dimensions a Chiral Magnetic term survives in the non-relativistic limit. In $2+1$ dimensions there is an anomalous Hall effect in inhomogeneous configurations, i.e.~a Hall current in the absence of magnetic field.

The paper is organized as follows: in \S~\ref{sec:hydro} we review Lifshitz hydrodynamics and derive the constitutive relations in parity breaking theories. We constrain possible terms to first order in derivatives using the local second law of thermodynamics. This is done for theories without Lorenz boost invariance, but otherwise relativistic.  In \S~\ref{sec:nr} we take a non-relativistic limit and derive the constitutive relations for theories without Galilean boost invariance. In \S~\ref{sec:drude} we study some phenomenological consequences of broken boost invariance using the Drude model of non-relativistic hydrodynamics with a drag term. Finally in \S~\ref{sec:conc} we summarize and discuss our results. We have collected several technical results in an Appendix~\ref{sec:app}.

\section{Lifshitz hydrodynamics with parity breaking terms}\label{sec:hydro}

The lack of invariance under Lorentz boosts in theories with Lifshitz symmetry implies that the energy-momentum tensor is not necessarily symmetric. Using rotational invariance one can show that no asymmetric terms are expected in the hydrodynamic energy-momentum tensor at the ideal level, and we confirmed this explicitly by a calculation in free field theory \cite{Hoyos:2013qna}. Note that rotational invariance is unbroken in the rest frame of the fluid. For an observer that sees the fluid moving with a velocity $u^\mu$ the corresponding symmetry condition is
\begin{equation}\label{rotinv}
(T^{\mu\nu}-T^{\nu\mu})P_{\mu\alpha}P_{\nu\beta}=0,
\end{equation}
where $P^{\mu\nu}=\eta^{\mu\nu}+u^\mu u^\nu$ is the projector in the directions transverse to the velocity. The absence of invariance under Lorentz boosts on the other hand allows the following components to be asymmetric: 
\begin{equation}
(T^{\mu\nu}-T^{\nu\mu})u_\mu P_{\nu\beta}\neq 0.
\end{equation}
In the rest frame of the fluid this becomes
\begin{equation}
T^{0i}-T^{i0}\neq 0.
\end{equation}
But in a state with rotational invariance the expectation value of $T^{0i}$ vanishes, so at the ideal order the hydrodynamic energy-momentum is symmetric. When the velocity or other fluid variables are not constant there can be asymmetric terms that depend on the derivatives of the fluid variables. In previous works \cite{Hoyos:2013eza,Chapman:2014hja} we found the possible asymmetric terms to first viscous order in fluids and superfluids with unbroken parity symmetry.  Here we will give a detailed presentation for charged fluids with parity breaking terms.

We will work in the Landau frame $T^{\mu\nu}u_\nu=-\varepsilon u^\mu$, where the energy-momentum tensor takes the form\footnote{We follow the usual bracket convention for symmetrized $A^{(\mu}B^{\nu)} = \frac{1}{2}(A^\mu B^\nu + A^\nu B^\mu)$, and antisymmetrized $A^{[\mu}B^{\nu]} = \frac{1}{2}(A^\mu B^\nu - A^\nu B^\mu)$ tensors.}
\begin{equation}\label{stress}
T^{\mu\nu}=\varepsilon u^\mu u^\nu +p P^{\mu\nu}+\pi_S^{(\mu\nu)}+\pi_A^{[\mu\nu]} +(u^\mu\pi_A^{[\nu\sigma]}+u^\nu\pi_A^{[\mu\sigma]})u_\sigma.
\end{equation}
Both $\pi_A$ and $\pi_S$ contain terms that are at least first order in derivatives. 
The last symmetric term depending on $\pi_A$ ensures that the Landau condition will be satisfied when we add the antisymmetric part. In addition, the symmetric part should be transverse $\pi^{(\mu\nu)}_S u_\nu=0$.  The condition \eqref{rotinv} constrains the anisymmetric part to be of the form
\begin{equation}
\pi_A^{[\mu\nu]}=u^{[\mu} V_A^{\nu]}.
\end{equation}
Where $V_A^\nu u_\nu=0$. Then, the form of the energy-momentum tensor in the Landau frame becomes
\begin{equation}\label{stress2}
T^{\mu\nu}=\varepsilon u^\mu u^\nu +p P^{\mu\nu}+\pi_S^{(\mu\nu)}+u^\mu V_A^\nu.
\end{equation}
In addition to the energy-momentum tensor, there is a conserved global current whose constitutive relation is
\begin{equation}\label{current}
J^\mu=q u^\mu +\nu^\mu,
\end{equation}
where we impose the condition $\nu^\mu u_\mu=0$, so that the charge density is $q=-J^\mu u_\mu$. 

The name of the game is to find all possible first order viscous terms, constructed from derivatives of the velocity, temperature $T$, chemical potential $\mu$ and external electromagnetic fields $A_\mu$. We consider the derivatives of the external fields to be of first order. There are many such terms that one can build, but we will constrain them by demanding a local form of the second law of thermodynamics holds. This means that it should be possible to define an entropy current (which coincides with the usual notion of entropy at equilibrium) whose divergence is not negative
\begin{equation}
\partial_\mu j_s^\mu\geq 0.
\end{equation}
Starting with Landau, this approach has been used with great success to constrain hydrodynamic coefficients. For parity breaking terms we will restrict the analysis to $2+1$ dimensions and $3+1$ dimensions. If parity is unbroken the entropy current takes the generic form
\begin{equation}
j_s^\mu = s u^\mu -\frac{\mu}{T}\nu^\mu.
\end{equation}
Where $s$ is the canonical entropy density $\varepsilon+p=Ts+\mu q$.
When parity is broken it is in principle possible to add to the entropy current additional terms that are pseudovectors, we will denote them as $\tv_a^\mu$. The entropy current  is then defined as
\begin{equation}
j_s^\mu= s u^\mu -\frac{\mu}{T}\nu^\mu+D_a \tv_a^\mu.
\end{equation}
We can derive the equation for the entropy current from the conservation equations
\begin{equation}\label{enteq}
\partial_\mu T^{\mu\nu}u_\nu+\mu \partial_\mu J^\mu=F^{\mu\nu}u_\mu J_\nu +\mu\cA.
\end{equation}
Here we allow for an anomaly $\partial_\mu J^\mu=\cA$ in the global current, such term can be present in $3+1$ dimensions. In $2+1$ dimension one should simply set $\cA=0$.  Expanding explicitly the equation above one finds
\begin{equation}
-T\partial_\mu(s u^\mu)+\partial_\mu \pi_S^{(\mu\nu)}u_\nu+u^\mu\partial_\mu V_A^\nu u_\nu+\mu \partial_\mu \nu^\mu=-E_\mu \nu^\mu+\mu \cA.
\end{equation}
Where we have used $\varepsilon+p=Ts+\mu q$, the first law
\begin{equation}
\partial_\mu\varepsilon=T \partial_\mu s+\mu \partial_\mu q,
\end{equation}
and we have defined the electric field as $E^\mu=F^{\mu\nu}u_\nu$. Straightforward algebra leads to the following equation for the entropy current
\begin{equation}\label{js}
\partial_\mu j_s^\mu=- \frac{1}{T}\pi_S^{(\mu\nu)}\partial_\mu u_\nu- \frac{1}{T}V_A^\mu a_\mu+\frac{1}{T} \nu^\mu \cE_\mu -\frac{\mu}{T} \cA+\partial_\mu (D_a \tv_a^\mu).
\end{equation}
Where $a^\mu=u^\alpha\partial_\alpha u^\mu$ is the acceleration and we have defined the vector
\begin{equation}
\cE_\mu=  E_\mu -TP_\mu^{\ \nu}\partial_\nu \left( \frac{\mu}{T}\right).
\end{equation}
The positivity condition $\partial_\mu j_s^\mu \geq 0$ implies that terms on the right hand side of the equation that do not cancel out should form squares. We can separate parity preserving and parity breaking contributions
\begin{equation}
\partial_\mu j_s^\mu=\Delta_{P}+\Delta_{\Pbr}.
\end{equation}
They do not mix, so we should impose the positivity condition separately for each of them
\begin{equation}
\Delta_{P}\geq 0,\ \ \Delta_{\Pbr}\geq 0.
\end{equation}
For illustration purposes we will review the analysis of \cite{Chapman:2014hja} for the parity preserving contributions. The most general form of the viscous terms allowed by the entropy current equation are
\begin{equation}
\begin{split}
\pi_S^{(\mu\nu)}=&-\eta \sigma^{\mu\nu}-\zeta P^{\mu\nu}\partial_\alpha u^\alpha,\\
V_A^\mu= & -\alpha_1 a^\mu-2\alpha_2 \cE^\mu,\\
\nu^\mu=&  2\alpha_3 a^\mu+\sigma \cE^\mu .
\end{split}
\end{equation}
Where $\eta$ and $\zeta$ are the shear and bulk viscosities respectively, $\sigma$ is the conductivity, and the shear tensor in $d+1$ dimensions is defined as
\begin{equation}
\sigma^{\mu\nu}=P^{\mu\alpha}P^{\nu\beta}\left(2\partial_{(\alpha} u_{\beta)}-\frac{2}{d}P_{\alpha\beta}(\partial_\sigma u^\sigma)\right).
\end{equation}
From \eqref{js}, the positivity condition imposes the following constraints on the parity preserving coefficients
\begin{equation}
\eta\geq 0,\ \zeta \geq 0,\ \sigma\geq 0,\ \ \sigma  \alpha_1 \geq (\alpha_2+\alpha_3)^2.
\end{equation}
The combination $\alpha_2-\alpha_3$ drops from the divergence, so it is dissipationless. It will vanish if we impose Onsager's relations. 

We will now proceed with the parity breaking terms. The analysis depends on the number of dimensions, so we treat them separately.

\subsection{Parity breaking transport in $3+1$ dimensions}

In $3+1$ dimensions there can be interesting new effects when parity is broken due to chiral anomalies. A purely hydrodynamic analysis shows that the value of some transport coefficients is fixed by the anomaly \cite{Son:2009tf,Sadofyev:2010pr,Neiman:2010zi}

There are two possible independent pseudovectors we can construct \cite{Son:2009tf}, the vorticity $\omega^\mu$ and the magnetic field $B^\mu$. They are defined as follows:
\begin{equation}
\begin{split}
\omega^\mu &= \frac{1}{2}\epsilon^{\mu \nu \rho \sigma}u_\nu \partial_\rho u_\sigma ,\\
B^\mu &= \frac{1}{2}\epsilon^{\mu \nu \rho \sigma}u_\nu F_{\rho\sigma}. 
\end{split}
\end{equation}
The entropy current has then the form
\begin{equation}
j_s^\mu=s u^\mu -\frac{\mu}{T}\nu^\mu+D_\omega \omega^\mu+D_BB^\mu.
\end{equation}
The parity breaking contribution to the divergence is
\begin{equation}
\Delta_{\Pbr}=-\frac{1}{T}V_{A,\Pbr}^\mu a_\mu+\frac{1}{T}\nu_{\Pbr}^\mu\cE_\mu+\partial_\mu(D_\omega \omega^\mu)+\partial_\mu(D_B B^\mu)-\frac{\mu}{T}C E_\mu B^\mu.
\end{equation}
Where we have introduced a possible chiral anomaly for the current $\cA=CE_\mu B^\mu$. Note that there are no $\omega^2$ or $B^2$, so the positivity condition for the entropy current is that there should be an exact cancellation among the parity breaking contributions 
\begin{equation}
\Delta_{\Pbr}=0.
\end{equation}
This implies that all the related transport is non-dissipative.  The divergence of the vorticity and of the magnetic field are not independent scalar quantities, but they can be decomposed as follows:
\begin{equation}
\partial_\mu \omega^\mu =2 a_\mu \omega^\mu, \ \ \partial_\mu B^\mu= a_\mu B^\mu-2\omega_\mu E^\mu.
\end{equation}
The possible parity breaking contributions that can appear in the viscous terms are
\begin{equation}
\begin{split}
V_{A,\, \Pbr}^\mu=& -T\beta_\omega \omega^\mu  - T\beta_B B^\mu,\\
\label{nua} \nu_\Pbr^\mu=&  \; \xi_\omega \omega^\mu + \xi_B B^\mu .
\end{split}
\end{equation}
It is sufficient if the cancellation is realized on-shell, this means that at this order we are allowed to use the hydrodynamic equations truncated to ideal order.  In particular, this fixes the acceleration 
\begin{equation}\label{acc}
a^\mu=\frac{q}{\varepsilon +p} E^\mu -\frac{1}{\varepsilon+p}P^{\mu\nu}\partial_\nu p.
\end{equation}
Collecting terms, we find 
\begin{align}\label{div3}
\Delta_{\Pbr} &= \omega^\mu \left( \partial_\mu D_\omega - \left(2D_\omega + \beta_\omega\right)\frac{ \partial_\mu p}{\varepsilon + p} - \xi_\omega \partial_\mu  \left( \frac{\mu}{T} \right) \right) \nonumber \\
&+ \omega^\mu E_\mu \left(\left(2D_\omega + \beta_\omega\right)\frac{q}{\varepsilon + p} - 2D_B + \frac{\xi_\omega}{T}\right) \nonumber \\
&+ B^\mu \left(\partial_\mu D_B - \frac{\partial_\mu p}{\varepsilon + p}\left(D_B + \beta_B\right) - \xi_B \partial_\mu\frac{\mu}{T}\right) \nonumber \\
&+B^\mu E_\mu \left(\frac{q \left(D_B + \beta_B\right)}{\varepsilon + p} + \frac{\xi_B}{T} - C\frac{\mu}{T}\right)= 0.
\end{align}
Each of the brackets has to vanish independently, this results in the following conditions: 
\begin{align}
&\partial_\mu D_\omega - \left(2D_\omega + \beta_\omega\right)\frac{ \partial_\mu p}{\varepsilon + p} - \xi_\omega \partial_\mu  \left( \frac{\mu}{T} \right)=0,\\
&\left(2D_\omega + \beta_\omega\right)\frac{q}{\varepsilon + p} - 2D_B + \frac{\xi_\omega}{T} = 0, \\
&\partial_\mu D_B - \frac{\partial_\mu p}{\varepsilon + p}\left(D_B + \beta_B\right) - \xi_B \partial_\mu\frac{\mu}{T} = 0, \\
&\frac{q \left(D_B + \beta_B\right)}{\varepsilon + p} + \frac{\xi_B}{T} - C\frac{\mu}{T} = 0. \label{The equations 4}
\end{align}
The derivation of the solutions can be found in the Appendix~\ref{app31}, here we present the final result:
\begin{align}
\beta_\omega(\bar{\mu},T) &=2\bar{\mu}T\beta_B+b_\omega,\\
\xi_B(\bar{\mu},T) & = C\left(\mu - \frac{1}{2}\frac{q \mu^2}{\varepsilon + p}\right) -\frac{q T}{\varepsilon + p}(d_B+\beta_B),\\
\xi_\omega(\bar{\mu},T) & =  C\left(\mu^2 - \frac{2}{3}\frac{q \mu^3}{\varepsilon + p}\right) +2T \left(d_B-\frac{2\mu q}{\varepsilon+p}d_B\right) \nonumber \\
&- \frac{q T}{\varepsilon+p}\left(2 d_\omega+2T\bar{\mu}\beta_B+b_\omega\right),\\
D_B(\bar{\mu},T) &=\frac{1}{2}CT\bar{\mu}^2+d_B,\\
D_\omega(\bar{\mu},T) &= \frac{1}{3}CT^2\bar{\mu}^3+2\bar{\mu}Td_B+d_\omega,\\
\label{dBin} d_B(T) &=\gamma_B T +T\int^T \frac{\beta_B(\tau)}{\tau^2}d\tau,\\
\label{dwin} d_\omega(T) &=\gamma_\omega T^2+T^2\int^T \frac{b_\omega(\tau)}{\tau^3}d\tau.
\end{align}
With $b_\omega(T)$ and $\beta_B(T)$ arbitrary functions of the temperature.

For $z=1$, we would expect $\beta_B=c_B T$ and $b_\omega=c_\omega T^2$ if there is scale invariance, as in theories with Lifshitz symmetry. However scale invariance would actually be broken because for this temperature dependence when we do the integrals over temperature in \eqref{dBin}, \eqref{dwin} we find logarithmic terms 
\begin{equation}
d_B=\gamma_B T+c_B T \log(T/\Lambda), \ \ d_\omega=\gamma_\omega T^2 +c_\omega T^2 \log(T/\Lambda),
\end{equation}
where $\Lambda$ is the scale associated to the breaking of symmetry. Therefore, Lifshitz symmetry demands $\beta_B=b_\omega=0$. In this case the transport coefficients are simplified to
\begin{align}
\beta_\omega(\bar{\mu},T) &=0,\\
\xi_B(\bar{\mu},T) & = C\left(\mu - \frac{1}{2}\frac{q \mu^2}{\varepsilon + p}\right) -\frac{q T^2}{\varepsilon + p}\gamma_B,\\
\xi_\omega(\bar{\mu},T) & =  C\left(\mu^2 - \frac{2}{3}\frac{q \mu^3}{\varepsilon + p}\right) +2\gamma_B T^2 -\frac{2 q}{\varepsilon+p}\left( 2\gamma_B \mu T^2 + \gamma_\omega T^3 \right),\\
D_B(\bar{\mu},T) &=\frac{1}{2}CT\bar{\mu}^2+\gamma_B T,\\
D_\omega(\bar{\mu},T) &= \frac{1}{3}CT^2\bar{\mu}^3+2\bar{\mu}T^2 \gamma_B+\gamma_\omega T^2.
\end{align}
Note that these are the same results as for the relativistic theory \cite{Newman:2005hd}. 
If we further impose CPT invariance (note that for a Lifshitz theory this is not a requirement), we will fix $\gamma_\omega=0$. For a detailed explanation on how the hydrodynamic quantities transform under CPT see \cite{Bhattacharya:2011tra}.

For $z \neq 1$, we would expect $\beta_B=c_B T^\frac{2-z}{z}$ and $b_\omega=c_\omega T^\frac{2}{z}$ if there is scale invariance, as in theories with Lifshitz symmetry. However, in this case the terms proportional to $\gamma_B$ and $\gamma_\omega$ break scale invariance. In this case, Lifshitz symmetry demands $\gamma_B=\gamma_\omega=0$. Therefore, the transport coefficients are 
\begin{align}
\beta_\omega(\bar{\mu},T) &=\left(2c_B\bar{\mu}+c_\omega\right)T^\frac{2}{z},\\
\xi_B(\bar{\mu},T) & = C\left(\mu - \frac{1}{2}\frac{q \mu^2}{\varepsilon + p}\right) -c_B\frac{2-z}{2(1-z)}\frac{q}{\varepsilon + p}T^\frac{2}{z},\\
\xi_\omega(\bar{\mu},T) & =  C\left(\mu^2 - \frac{2}{3}\frac{q \mu^3}{\varepsilon + p}\right) +c_B\frac{z}{1-z}\left(1-\frac{2\mu q}{\varepsilon + p}\right)T^\frac{2}{z} \nonumber \\
&-\frac{q}{\varepsilon+p}\left( \frac{c_\omega}{1-z}+2c_B \bar{\mu}\right)T^\frac{2+z}{z},\\
D_B(\bar{\mu},T) &=\frac{1}{2}CT\bar{\mu}^2+\frac{z}{2(1-z)}c_B T^\frac{2-z}{z},\\
D_\omega(\bar{\mu},T) &= \frac{1}{3}CT^2\bar{\mu}^3+\frac{z}{2(1-z)}\left(2\bar{\mu}c_B+c_\omega\right)T^\frac{2}{z}.
\end{align}

\subsection{Parity breaking transport in $2+1$ dimensions}

We follow closely the analysis of parity breaking fluids in \cite{Jensen:2011xb}. There are three possible independent transverse pseudovectors we can construct
\begin{equation}\label{Udef}
\begin{split}
\tilde{U}_1^\mu =& \epsilon^{\mu \nu \rho}u_\nu a_\rho ,\\
\tilde{U}_2^\mu=& \epsilon^{\mu \nu \rho}u_\nu E_\rho,\\
\tilde{U}_3^\mu =& -\frac{1}{T}\epsilon^{\mu \nu \rho}u_\nu \cE_\rho.
\end{split}
\end{equation}
We can also construct two independent peudoscalars, the vorticity $\omega$ and the magnetic field $B$, and a pseudotensor, the Hall viscosity $\tilde{\sigma}^{\mu\nu}$: \footnote{An useful relation is $2\partial_{[\mu} u_{\nu]}=-2\epsilon_{\mu\nu\rho} u^\rho \omega$.}
\begin{equation}
\begin{split}
\omega =& \frac{1}{2}\epsilon^{\mu \nu \rho}u_\mu \partial_\nu u_\rho ,\\
B =& \frac{1}{2}\epsilon^{\mu \nu \rho}u_\mu F_{\nu\rho},\\
\tilde{\sigma}^{\mu\nu}= &\frac{1}{4}\left(\epsilon^{\mu\rho\sigma}u_\rho \sigma_\sigma^{\ \nu}+\epsilon^{\nu\rho\sigma}u_\rho \sigma_\sigma^{\ \mu} \right).
\end{split}
\end{equation}
Note that our definitions differ from those of \cite{Jensen:2011xb}, the relation is $B_{\rm theirs}=-B_{\rm ours}$ and $\Omega_{\rm theirs}=-2\omega_{\rm ours}$. It will be convenient for the analysis of the entropy current to group all the possible pseudovectors (transverse and not transverse) in the following basis ($R=\frac{q}{\varepsilon+p}$)
\begin{equation}\label{Vdef}
\begin{split}
\tilde{V}_1^\mu =& \epsilon^{\mu \nu \rho}u_\nu \partial_\rho T=-T \tilde{U}_1^\mu-R T^2 \tilde{U}_3^\mu ,\\
\tilde{V}_2^\mu=& \tilde{U}_2^\mu,\\
\tilde{V}_3^\mu =& \epsilon^{\mu \nu \rho}u_\nu \partial_\rho \frac{\mu}{T}=\tilde{U}_3^\mu+\frac{1}{T}\tilde{U}_2^\mu,\\
\tilde{V}_4^\mu=& \frac{1}{2}\epsilon^{\mu\nu\rho}F_{\nu\rho}=\tilde{U}_2^\mu-B u^\mu,\\
\tilde{V}_5^\mu =& \epsilon^{\mu\nu\rho}\partial_\nu u_\rho=-\tilde{U}_1^\mu-2\omega u^\mu.
\end{split}
\end{equation}

We can add the following parity breaking terms  in the viscous contributions to the energy-momentum tensor and the currents:
\begin{itemize}
\item Symmetric part of the energy-momentum tensor:
\begin{equation}
\pi_{S,\Pbr}^{(\mu\nu)}=-\zeta_\omega P^{\mu\nu}\omega-\zeta_B P^{\mu\nu} B-\eta_H \tilde{\sigma}^{\mu\nu}.
\end{equation}
\item Asymmetric part of the energy-momentum tensor:
\begin{equation}
\begin{split}
V_{A,\Pbr}^\mu=& -T\sum_{i=1}^3\tilde{\mu}_i\tilde{V}_i^\mu.
\end{split}
\end{equation}
\item Current:
\begin{equation}
\label{nua2} \nu_{\Pbr}^\mu= \sum_{i=1}^3\tilde{\delta}_i\tilde{V}_i^\mu.
\end{equation}
\item Entropy current
\begin{equation}
j_s^\mu =s u^\mu -\frac{\mu}{T}\nu^\mu +\sum_{i=1}^5 \tilde{\nu}_i \tilde{V}_i^\mu  .
\end{equation}
\end{itemize}
The parity breaking contribution to the divergence of the entropy current is
\begin{align}
\Delta_{\Pbr}&=-\frac{1}{T}\pi^{(\mu\nu)}_{S,\Pbr}\partial_\mu u_\nu-\frac{1}{T}V_{A,\Pbr}^\mu U^{(1)}_\mu-\nu_{\Pbr}^\mu U^{(3)}_\mu+\sum_{i=1}^5\left(\frac{\partial\tilde{\nu}_i}{\partial T}\partial_\mu T \tilde{V}^\mu_i+\frac{\partial\tilde{\nu}_i}{\partial \bar{\mu}}\partial_\mu \bar{\mu} \tilde{V}^\mu_i+\tilde{\nu}_i\partial_\mu\tilde{V}^\mu_i\right)
\end{align}
Where we have set the anomalous term to zero $\cA=0$ since the number of spacetime dimensions is odd. 
The positivity condition for the entropy current is that there should be an exact cancellation among the parity breaking contributions, since it is not possible to form squares
\begin{equation}
\Delta_{\Pbr}=0.
\end{equation}
This implies that all the related transport is non-dissipative.  The divergences of the pseudovectors  are not independent scalar quantities, but they can be decomposed as follows:
\begin{align}
&\partial_\mu \tilde{V}^\mu_1 = -2\omega u^\alpha\partial_\alpha T - \tilde{U}^\mu_1\partial_\mu T, \\
&\partial_\mu \tilde{V}^\mu_2 = \partial_\alpha (B u^\alpha), \\
&\partial_\mu \tilde{V}^\mu_3 = -2\omega u^\alpha\partial_\alpha \frac{\mu}{T} - \tilde{U}^\mu_1\partial_\mu \frac{\mu}{T},\\
&\partial_\mu \tilde{V}^\mu_4=0,\\
&\partial_\mu \tilde{V}^\mu_5=0.
\end{align}
It is sufficient if the cancellation is realized on-shell, this means that at this order we are allowed to use the hydrodynamic equations truncated to ideal order.  In particular, this fixes the acceleration to \eqref{acc} and we have the following relations among derivatives
\begin{equation}
\begin{split}
u^\mu \partial_\mu T=& -T \left(\frac{\partial p}{\partial \epsilon}\right)_{q} \theta,\\
u^\mu \partial_\mu \bar{\mu}=& -\frac{1}{T} \left(\frac{\partial p}{\partial q}\right)_{\varepsilon} \theta,\\
u^\mu \partial_\mu p=& -(\varepsilon+p)\left(\left(\frac{\partial p}{\partial \epsilon}\right)_{q} +R \left(\frac{\partial p}{\partial q}\right)_{\varepsilon}  \right)\theta,\\
P^{\mu\alpha}\partial_\alpha T =& -TU_1^\mu-RT^2 U_3^\mu =-Ta^\mu +RT \cE^\mu,\\
P^{\mu\alpha}\partial_\alpha\bar{\mu}=& U_3^\mu +\frac{1}{T} U_2^\mu =\frac{1}{T}(-\cE^\mu+E^\mu),\\
P^{\mu\alpha}\partial_\alpha p=& -(\varepsilon+p)(a^\mu -R E^\mu).\\
\end{split}
\end{equation}
Collecting terms, we find 
\begin{align} \label{div2}
\Delta_{\Pbr} &=U_1 \tilde{U_3}\left(\tilde{\mu}_3-R T^2\tilde{\mu}_1-T\tilde{\delta}_1 -R T^2(\partial_T\tilde{\nu}_5+\tilde{\nu}_1)+(\partial_{\bar{\mu}} \tilde{\nu}_5+\tilde{\nu}_3)+T(\partial_{\bar{\mu}} \tilde{\nu}_1-\partial_T \tilde{\nu}_3)\right) \nonumber \\
& + U_1 \tilde{U_2}\left(\tilde{\mu}_2+\frac{1}{T}\tilde{\mu}_3 + \frac{\partial_{\bar{\mu}} \tilde{\nu}_5+\tilde{\nu}_3}{T} + \partial_{\bar{\mu}} \tilde{\nu}_1 - \partial_T \tilde{\nu}_3-T\partial_T \tilde{\nu}_4\right) \nonumber \\
&  + U_2 \tilde{U_3}\left(\tilde{\delta}_2+\frac{1}{T}\tilde{\delta}_3 + R T (\partial_T \tilde{\nu}_3 -\partial_{\bar{\mu}} \tilde{\nu}_1) - \partial_{\bar{\mu}} \tilde{\nu}_4 + R T^2 \partial_T \tilde{\nu}_4 \right) \nonumber \\
& + B \theta \left(\frac{1}{T}\zeta_B+T\left(\frac{\partial p}{\partial \epsilon}\right)_{q} \partial_T \tilde{\nu}_4 + \frac{1}{T}\left(\frac{\partial p}{\partial q}\right)_{\epsilon}\partial_{\bar{\mu}}\tilde{\nu}_4\right) \nonumber \\
& + 2\omega\theta\left(\frac{1}{2T}\zeta_\omega + T\left(\frac{\partial p}{\partial \epsilon}\right)_{q} (\partial_T \tilde{\nu}_5 + \tilde{\nu}_1) + \frac{1}{T}\left(\frac{\partial p}{\partial q}\right)_{\epsilon}(\partial_{\bar{\mu}} \tilde{\nu}_5 + \tilde{\nu}_3 )\right) \nonumber \\
& + \partial_\mu\tilde{\nu}_2 \tilde{U_2}  + \tilde{\nu}_2 B \theta + \tilde{\nu}_2 u^\mu\partial_\mu B
\end{align}
Each of the brackets has to vanish independently, this results in the following conditions: 
\begin{align}
&\tilde{\nu}_2 = 0 \\
&\tilde{\mu}_3-R T^2\tilde{\mu}_1-T\tilde{\delta}_1 -R T^2(\partial_T\tilde{\nu}_5+\tilde{\nu}_1)+(\partial_{\bar{\mu}} \tilde{\nu}_5+\tilde{\nu}_3)+T(\partial_{\bar{\mu}} \tilde{\nu}_1-\partial_T \tilde{\nu}_3)=0 \\
&\tilde{\mu}_2+\frac{1}{T}\tilde{\mu}_3 + \frac{\partial_{\bar{\mu}} \tilde{\nu}_5+\tilde{\nu}_3}{T} + \partial_{\bar{\mu}} \tilde{\nu}_1 - \partial_T \tilde{\nu}_3-T\partial_T \tilde{\nu}_4 = 0\\
&\tilde{\delta}_2+\frac{1}{T}\tilde{\delta}_3 + R T (\partial_T \tilde{\nu}_3 -\partial_{\bar{\mu}} \tilde{\nu}_1) - \partial_{\bar{\mu}} \tilde{\nu}_4 + R T^2 \partial_T \tilde{\nu}_4   = 0 \\
&\frac{1}{T}\zeta_B+T\left(\frac{\partial p}{\partial \epsilon}\right)_{q} \partial_T \tilde{\nu}_4 + \frac{1}{T}\left(\frac{\partial p}{\partial q}\right)_{\epsilon}\partial_{\bar{\mu}}\tilde{\nu}_4 = 0 \\
&\frac{1}{2T}\zeta_\omega + T\left(\frac{\partial p}{\partial \epsilon}\right)_{q} (\partial_T \tilde{\nu}_5 + \tilde{\nu}_1) + \frac{1}{T}\left(\frac{\partial p}{\partial q}\right)_{\epsilon}(\partial_{\bar{\mu}} \tilde{\nu}_5 + \tilde{\nu}_3 ) = 0
\end{align}
For $\tilde{\mu}_i=0$ we recover the results of \cite{Jensen:2011xb}.
If we assume Lifshitz scale invariance, in general the coefficientes will have a  power  dependence on the temperature times an arbitrary function of $\bar{\mu}$. The power is determined by demanding that all the terms in the same equation have the same scaling, taking into account that the scaling dimension of the temperature is $[T]=z$. This fixes:  
\begin{equation}
\begin{split}
&\tilde\mu_1 \sim T^{\frac{1-z}{z}}, \ \ \tilde\mu_2 \sim T^{\frac{1-z}{z}}, \ \ \tilde\mu_3 \sim T^{\frac{1}{z}}, \\
&\tilde\delta_1 \sim T^{\frac{1-z}{z}}, \ \ \tilde\delta_2 \sim T^{\frac{1-z}{z}}, \ \ \tilde\delta_3 \sim T^{\frac{1}{z}}, \\
&\tilde\nu_1 \sim T^{\frac{1-z}{z}}, \ \ \tilde\nu_3  \sim T^{\frac{1}{z}}, \ \ \tilde\nu_4 \sim T^{\frac{1-z}{z}}, \ \ \tilde\nu_5 \sim T^{\frac{1}{z}},\\
&\zeta_B \sim T^{1/z}, \ \ \zeta_\omega \sim T^{\frac{z+1}{z}}
\end{split}
\end{equation}

\section{The non-relativistic limit}\label{sec:nr}

We now study fluids with broken Galilean boost invariance. Instead of deriving the hydrodynamic constitutive relations from the start, we will take a non-relativistic limit of the Lifshitz hydrodynamic equations we have discussed in the previous sections. This will automatically lead to non-relativistic hydrodynamic equations. Similar limits have been used in the context of Galilean invariant fluids in \cite{Kaminski:2014,Jensen:2014wha}.

In the relativistic fluid the maximal velocity $c$ appears in $u^\mu = \left(1,\beta\right)/\sqrt{1-\beta^2}$, where $\beta^i = v^i/c$. In the non-relativistic limit $c\rightarrow \infty$, the pressure is not affected while the relativistic energy is expanded in terms of the mass density $\rho$ and the internal energy $U$ as 
\begin{align}
\varepsilon=c^2 \rho - \frac{\rho v^2}{2}+U
\end{align}
From the thermodynamic relation $\varepsilon+p=Ts+\mu q$, we get that
\begin{equation}\label{cexpq}
	q=\rho c-\rho\frac{v^2}{2c}, \ \ \mu=c+\frac{\mu_{NR}}{c}
\end{equation}
The electromagnetic fields scale as $A_i\to A_i$, $A_0\to A_t/c$, which give the relations: \\
In 3+1 dimensions:
\begin{align}
E^i=\frac{1}{c}\left(E^i_{NR}+\epsilon^{ijk}v_jB_{NR,k}\right) + o(1/c), \  \ B^i = B_{NR}^i + O(1/c^2).
\end{align}
In 2+1 dimensions:
\begin{align}
E^i=\frac{1}{c}\left(E^i_{NR}-\epsilon^{ij}v_jB_{NR}\right) + o(1/c), \ \ B = B_{NR} + O(1/c^2).
\end{align}
The subscript $NR$ will be left out in the reminder of the text. We use the conventions in which $\epsilon^{0123}=\epsilon^{123}=\epsilon^{12}=1$.

For charged hydrodynamics we first go to the Eckart frame, where $J^\mu=q u^\mu$. To first order in derivatives the energy-momentum tensor in this frame is
\begin{equation}
	T^{\mu\nu}=\varepsilon u^\mu u^\nu +p P^{\mu\nu}+u^\mu V_A^\nu +\pi^{\mu\nu}-\frac{\varepsilon+p}{q}(u^\mu \nu^\nu+u^\nu \nu^\mu).
\end{equation}
Note that in the Eckart frame the current conservation equation gives the usual continuity equation. If $\nu^i\sim O(1)$, the energy conservation equation becomes
\begin{equation}
	\partial_t \rho+\partial_i(\rho v^i-\nu^i)=0.
\end{equation}
This would make the charge and mass currents different, while we are interested in describing systems made of particles with fixed charge over mass ratio.  In order to  have a consistent expansion, the dissipative terms in $\nu^i$ should start contributing only to the next order, where they enter in the thermal current. Then we have to impose that  $\nu^i\sim O(1/c^2)$. In this case, there are no contributions from $\nu^i$ terms to the Navier-Stokes equations. 

Galilean boost invariance will fix a relation between momentum density and mass current $P^i=J^i=\rho v^i$. When we take the non-relativistic limit of the boost-breaking hydrodynamics, this condition does not hold anymore beyond ideal order. We find that the Navier-Stokes equations become
\begin{itemize}
\item In $3+1$ dimensions:
\begin{equation}
\begin{split}
&\partial_t P^i+ \partial_j \left(P^i v^j\right) +\partial^ip  =\\
&\rho\left(E^i+\epsilon^{ijk}v_j B_k\right)+ \partial_j\left(\eta\sigma^{ij} + \delta^{ij}\zeta \partial_k v^k\right) .
\end{split}
\end{equation}
Where $\sigma_{ij}=\partial_i v_j +\partial_j v_i-\frac{2}{d}\delta_{ij}\partial_k v^k$ is the shear tensor (in this case for $d=3$).
The momentum density is 
\begin{equation}\label{P3d}
P^i=\rho v^i - \alpha_a a^i-\alpha_T \partial^i T-T\beta_B B^ i,
\end{equation}
where we define the acceleration as $a^i = D_t v^i=\partial_tv^i+v^j\partial_jv^i$ and the magnetic field as $B^ i = \frac{1}{2}\epsilon^{ijk}F_{jk}$. The term $\beta_B$ allows a Chiral Magnetic Effect in a non-relativistic theory.  If the theory was obtained from a non-relativistic limit, it is determined by the value of the anomaly in the original relativistic theory. Note that the continuity equation for the current holds in this limit, and the effect of the anomaly enters in the thermal current and the momentum density.

We got these expressions by demanding that the leading order dependence of the relativistic coefficients with the speed of light is,
\begin{align}
&\eta \sim O\left(c\right), \ \ \zeta \sim O\left(c\right), \ \ \alpha \sim O\left(c^3\right), \ \ \beta_\omega \sim O\left(c^2\right), \ \ \beta_B \sim O\left(c\right), \ \ \alpha_2 \sim O\left(1\right), \ \ \nonumber \\
&\alpha_3 \sim O\left(1\right), \ \ \sigma \sim O\left(\frac{1}{c^3}\right), \ \ \xi_\omega \sim O\left(\frac{1}{c}\right), \ \ \xi_B \sim O\left(\frac{1}{c^2}\right), \ \ C \sim O\left(\frac{1}{c}\right).
\end{align}
In principle we could add to the momentum a term of the form 
\begin{equation}
\beta_\omega \Omega^i,
\end{equation}
where $\Omega^i=\frac{1}{2}\epsilon^{ijk}\partial_k v_j$ is the vorticity. However, the entropy analysis that we detail in Appendix~\ref{entropy1} reveals that those terms should be zero in the non-relativistic limit. Therefore, Chiral Vortical Effects are suppressed in the non-relativistic limit.

\item In $2+1$ dimensions:
\begin{equation}
\begin{split}
&\partial_t P^i+ \partial_j \left(P^i v^j\right) +\partial^ip  =\\
&\rho\left(E^i-\epsilon^{ij}v_j B\right)+ \partial_j\left(\eta\sigma^{ij}+\eta_H\tilde{\sigma}^{ij} + \delta^{ij}\zeta\partial_k v^k\right) 
\end{split}
\end{equation}
Where the Hall viscosity tensor is
\begin{equation}
\tilde{\sigma}^{ij}=\frac{1}{2}(\epsilon^{ik}\sigma_k^j+\epsilon^{jk}\sigma_k^i),
\end{equation}
and $\sigma_{ij}$ is the shear tensor for $d=2$. The momentum density is\footnote{The coefficients $\beta$ are combinations of the original coefficients $\tilde{\mu}$ in the non-relativistic limit. We have separated the different independent contributions.} 
\begin{equation}\label{P2d}
P^i=\rho v^i - \alpha_a a^i-\alpha_T \partial^i T - \beta_T\epsilon^{ij}\partial_j T -\beta_\mu\epsilon^{ij}\partial_j \mu_{NR}-\beta_E\epsilon^{ij}(E_j-\epsilon_{jk}v^k B).
\end{equation}
We got this expression by demanding that the leading order dependence of the relativistic coefficients with the speed of light is,
\begin{align}
&\zeta_\omega \sim \eta_H \sim O(c), \ \ \zeta_B \sim O(1), \\
&\tilde\mu_1 = \frac{c^3}{T^2}\tilde\mu_3^{(0)} + c\tilde\mu_1^{(1)}+ O(c), \ \ \tilde\mu_3 = c^2\tilde\mu_3^{(0)} + \tilde\mu_3^{(1)} + O(1), \ \ \tilde\mu_2 \sim O(c^2),\\
&\tilde\delta_1 = \tilde\delta_1^{(0)} + O(1/c^2), \ \ \tilde\delta_3 = \frac{T^2}{c}\tilde\delta_1^{(0)} + O(1/c^3), \ \ \tilde\delta_2 \sim O(1/c).
\end{align}
In principle we could add to the stress tensor terms of the form 
\begin{equation}
\delta^{ij}(\zeta_B B+\zeta_\Omega \Omega),
\end{equation}
with $B$ the magnetic field and $\Omega=-\frac{1}{2}\epsilon^{ij}\partial_i v_j$ the vorticity. However, the entropy analysis that we detail in Appendix~\ref{entropy2} reveals that those terms should be zero in the non-relativistic limit.
\end{itemize}

\section{Drude Model}\label{sec:drude}

We model the collective motion of electrons in the strange metal as a charged fluid moving through a static medium, that produces a drag on the fluid.
The hydrodynamic equations are
\begin{align}
\partial_\mu J^\mu =\cA, \ \ \partial_\mu T^{\mu \nu} = F^{\nu\sigma} J_\sigma-\lambda  c \delta^{\nu i} J_i.
\end{align}
We are interested in describing a steady state where the fluid has been accelerated by the electric field, increasing the current until the drag force is large enough to compensate for it. In order for this to happen we keep the external fields constant in time, although we allow them to vary slowly in space. In the steady state configuration the current does not change but other quantities like the energy can change with time since there is dissipation. 

\subsection{Effects in 3+1 dimensions}

The Navier-Stokes equations with the drag term included are
\begin{equation}
\begin{split}
&\partial_t P^i+ \partial_j \left(P^i v^j\right) +\partial^ip  =\\
&\rho\left(E^i+\epsilon^{ijk}v_j B_k\right)+ \partial_j\left(\eta\sigma^{ij} + \delta^{ij}\zeta \partial_k v^k\right) -\lambda \rho v^i.
\end{split}
\end{equation}
Where $P^i$ is given by \eqref{P3d}. We can see the effect of breaking of boost invariance and parity by introducing constant electric and a slowly varying magnetic field $E_x$ and $B_z(x)$, in such a way that when we solve for the velocity derivative terms are suppressed.  We also assume that the amplitude of the electromagnetic field is small, so non-linear terms are subleading. To leading order, the current has only a longitudinal component
\begin{equation}
J_x =\frac{\rho}{\lambda}E_x.
\end{equation}
Since $\partial_i v^i=0$ to this order, we can take the density $\rho$ to be constant, as well as the temperature and the chemical potential.

In addition to subleading corrections to the longitudinal current, currents in the transverse directions are also generated. There are two kind of contributions, pointing in different directions. One is coming from the Lorenz force term
\begin{equation}
J_y = -\frac{\rho}{\lambda^2}E_x B_z.
\end{equation}
The second is due to the Chiral Magnetic term and points in the direction of the magnetic field
\begin{equation}
J_z = \frac{T\beta_B}{\lambda^2}E_x\partial_x B_z.
\end{equation}
Note that this current would be forbidden in a Galilean-invariant theory (with a single species of particles). 

\subsection{Effects in 2+1 dimensions}

Assuming a time-independent configuration, the hydrodynamic equations for the density and the velocity are
\begin{equation}
\partial_i(\rho v^i)=0,\ \ \partial_k (P^i v^k)+\partial^i p=\partial_k\left( \eta \sigma^{ki}+\eta_H \tilde{\sigma}^{ki}+\zeta \delta^{ki}\theta\right)-\lambda \rho v^i +\rho (E^i-\epsilon^{ij}v_j B).
\end{equation}
Where the momentum $P^i$, given by \eqref{P2d}, is  to first order
\begin{equation}
P^i=\rho v^i-\alpha_a v^k\partial_k v^i-\alpha_T \partial^i T-\beta_T \epsilon^{ij}\partial_j T-\beta_\mu \epsilon^{ij}\partial_j \mu_{NR}-\beta_E \epsilon^{ij} (E_j-\epsilon_j^{\ k}v_k B).
\end{equation}
We will allow a constant electric field $E_x$ and gradients of temperature and chemical potential along the same direction $T(x)$, $\mu(x)$. This implies that the pressure and the density are not constant but also vary along the $x$ direction $p(x)$, $\rho(x)$. We will assume that the gradients are small 
so we can do a derivative expansion and solve order by order around constant density and velocities.

To leading order we get the usual longitudinal current
\begin{equation}
J_x= \frac{\rho}{\lambda}E_x.
\end{equation}
A transverse (Hall) current is also generated. The leading contribution to the velocity in the $y$ direction is
\begin{equation}
J_y=\frac{1}{\lambda^2 }\left[\frac{\beta_E}{ \rho}\partial_x^2 p-\beta_T \partial^2_x T-\beta_\mu\partial_x^2 \mu_{NR} \right]E_x.
\end{equation}
This can be interpreted as an anomalous Hall effect: there is a transverse current in the absence of magnetic fields.

There can also be a higher order correction due to the Hall viscosity $O(\partial_x^3)$. Therefore, this non-homogeneous configuration allows to measure the new transport coefficients from a Hall current generated in the absence of magnetic fields. In the case of a fluid with Galilean invariance the Hall viscosity would be the leading order contribution. One can distinguish them in principle because the Lifshitz contributions would be largest close to a minimum or a maximum of the pressure, temperature and or chemical potential, while the contribution from the Hall viscosity would approximately vanish.

\section{Summary and conclusions}\label{sec:conc}

We have studied the constitutive relations of fluids for systems with Lifshitz symmetry and broken parity. We find new possible effects in the presence of magnetic field or vorticity. These are qualitatively different from theories with unbroken boost invariance and could be used to discern whether quantum critical points are really Lifshitz or a different type of theories with scaling symmetry. 

When the condition of boost invariance is relaxed  there can be new terms in the energy-momentum tensor that can be grouped in a vector  $V_A^\mu$ 
\begin{equation}
\begin{split}
T^{\mu\nu}=& \varepsilon u^\mu u^\nu +p P^{\mu\nu}+\pi_S^{(\mu\nu)}+u^\mu V_A^\nu,\\
J^\mu= &q u^\mu +\nu^\mu.
\end{split}
\end{equation}
In $3+1$ dimensions the terms that break parity are proportional to the magnetic field or the vorticity 
\begin{equation}
\begin{split}
V_{A,\, \Pbr}^\mu=& -T\beta_\omega \omega^\mu  - T\beta_B B^\mu,\\
\nu_\Pbr^\mu=&  \; \xi_\omega \omega^\mu + \xi_B B^\mu .
\end{split}
\end{equation}
If there was boost invariance, both $\xi_\omega$ and $\xi_B$ will be present only if there are chiral anomalies. In a Lifshitz theory this is not necessary, however in order to preserve the scaling symmetry for $z=1$, one should set the coefficients $\beta_\omega=\beta_B=0$. 

In $2+1$ dimensions there are many more terms allowed
\begin{equation}
\begin{split}
\pi_{S,\Pbr}^{(\mu\nu)}=& -\zeta_\omega P^{\mu\nu}\omega-\zeta_B P^{\mu\nu} B-\eta_H \tilde{\sigma}^{\mu\nu},\\
V_{A,\Pbr}^\mu=& -T\sum_{i=1}^3\tilde{\mu}_i\tilde{V}_i^\mu,\\
\nu_{\Pbr}^\mu=& \sum_{i=1}^3\tilde{\delta}_i\tilde{V}_i^\mu.
\end{split}
\end{equation}
Where the definition of the pseudovectors $\tilde{V}_i^\mu$ can be found in \eqref{Udef}, \eqref{Vdef}. They are constructed with epsilon tensors and the gradients of temperature and chemical potential and the acceleration and electric field. The entropy analysis imposes some constraints among the coefficients but all the terms are allowed.

In the non-relativistic limit the breaking of boost invariance can be seen as a modification of the momentum density, which is no longer equal to the current $J^i=\rho v ^i$ but receives corrections to first order. In $3+1$ dimensions there is a parity breaking term proportional to the magnetic field
\begin{equation}
P^i=\rho v^i - \alpha_a a^i-\alpha_T \partial^i T-T\beta_B B^ i.
\end{equation}
In this case the second law forbids a term proportional to the vorticity but the scaling symmetry does not impose  the condition that $\beta_B=0$. Therefore the relativistic and non-relativistic cases are qualitatively different. Using a Drude model with drag coefficient $\lambda$ for the strange metal, the parity breaking term is responsible for producing a current in the direction of the magnetic field, if the field is inhomogeneous
\begin{equation}
J_z = \left[\frac{T\beta_B}{\lambda^2}\partial_x B_z\right] E_x .
\end{equation}

In $2+1$ dimensions the momentum density receives several parity breaking contributions
\begin{equation}
P^i=\rho v^i-\alpha_a v^k\partial_k v^i-\alpha_T \partial^i T-\beta_T \epsilon^{ij}\partial_j T-\beta_\mu \epsilon^{ij}\partial_j \mu_{NR}-\beta_E \epsilon^{ij} (E_j-\epsilon_j^{\ k}v_k B).
\end{equation}
Using the Drude model, gradients of the temperature and/or chemical potential will introduce an anomalous Hall conductivity
\begin{equation}
J_y=\frac{1}{\lambda^2 }\left[\frac{\beta_E}{ \rho}\partial_x^2 p-\beta_T \partial^2_x T-\beta_\mu\partial_x^2 \mu_{NR} \right]E_x.
\end{equation}

\acknowledgments

C.H. thanks the Galileo Galilei Institute for Theoretical Physics for the hospitality and the INFN for partial support during the completion of this work.
This work is supported in part by the Israeli Science Foundation Center of Excellence, BSF, GIF and the I-CORE program of Planning and Budgeting Committee and the Israel Science Foundation (grant number 1937/12).  This work is partially supported by the Spanish grant MINECO-13-FPA2012-35043-C02-02. C.H. is supported by the Ramon y Cajal fellowship RYC-2012-10370. 

\appendix

\section{Entropy calculations}\label{sec:app}

\subsection{Positivity conditions in $3+1$ dimensions}\label{app31}

In order to solve the conditions
\begin{align}
&\partial_\mu D_\omega - \left(2D_\omega + \beta_\omega\right)\frac{ \partial_\mu p}{\varepsilon + p} - \xi_\omega \partial_\mu  \left( \frac{\mu}{T} \right)=0,\\
&\left(2D_\omega + \beta_\omega\right)\frac{q}{\varepsilon + p} - 2D_B + \frac{\xi_\omega}{T} = 0, \\
&\partial_\mu D_B - \frac{\partial_\mu p}{\varepsilon + p}\left(D_B + \beta_B\right) - \xi_B \partial_\mu\frac{\mu}{T} = 0, \\
&\frac{q \left(D_B + \beta_B\right)}{\varepsilon + p} + \frac{\xi_B}{T} - C\frac{\mu}{T} = 0.
\end{align}
We first take $\bar{\mu}=\mu/T$ and $p$ as independent thermodynamic variables, so that 
\begin{equation}
\partial_\mu D_{B,\omega}=\left(\frac{\partial D_{B,\omega}}{\partial \bar{\mu}}\right)_p \partial_\mu \bar{\mu}+\left(\frac{\partial D_{B,\omega}}{\partial p}\right)_{\bar{\mu}} \partial_\mu p.
\end{equation}
Collecting terms of the same type we get
\begin{align}
&\left(\frac{\partial D_{\omega}}{\partial p}\right)_{\bar{\mu}} - \frac{ 2D_\omega + \beta_\omega}{\varepsilon + p} =0,\\
&\left(\frac{\partial D_{\omega}}{\partial \bar{\mu}}\right)_p - \xi_\omega =0,\\
&\left(2D_\omega + \beta_\omega\right)\frac{q}{\varepsilon + p} - 2D_B + \frac{\xi_\omega}{T} = 0, \\
&\left(\frac{\partial D_{B}}{\partial p}\right)_{\bar{\mu}}  - \frac{D_B + \beta_B}{\varepsilon + p} = 0, \\
&\left(\frac{\partial D_{B}}{\partial \bar{\mu}}\right)_p- \xi_B  = 0, \\
&\frac{q \left(D_B + \beta_B\right)}{\varepsilon + p} + \frac{\xi_B}{T} - C\bar{\mu} = 0.
\end{align}
We can trade derivatives with respect to $p$ by derivatives with respect to $T$ by using the equations
\begin{align}
\left(\frac{\partial T}{\partial p}\right)_{\bar{\mu}} =  \frac{T}{\varepsilon + p},\hspace{30pt}
\left(\frac{\partial T}{\partial\bar{\mu}}\right)_p = - \frac{q T^2}{\varepsilon + p}
\end{align}
Then,
\begin{align}
&\label{domega} T\left(\frac{\partial D_{\omega}}{\partial T}\right)_{\bar{\mu}} - (2D_\omega + \beta_\omega) =0,\\
& \left(\frac{\partial D_{\omega}}{\partial \bar{\mu}}\right)_T-\frac{qT^2}{\varepsilon+p}\left(\frac{\partial D_{\omega}}{\partial T}\right)_{\bar{\mu}} - \xi_\omega =0,\\
&\left(2D_\omega + \beta_\omega\right)\frac{q}{\varepsilon + p} - 2D_B + \frac{\xi_\omega}{T} = 0, \\
&\label{dB} T\left(\frac{\partial D_{B}}{\partial T}\right)_{\bar{\mu}}  - (D_B + \beta_B) = 0, \\
& \left(\frac{\partial D_{B}}{\partial \bar{\mu}}\right)_T-\frac{qT^2}{\varepsilon+p}\left(\frac{\partial D_{B}}{\partial T}\right)_{\bar{\mu}} - \xi_B  = 0,\\
&\frac{q \left(D_B + \beta_B\right)}{\varepsilon + p} + \frac{\xi_B}{T} - C\bar{\mu} = 0.
\end{align}
Combining the first three and the last three equations we get
\begin{align}
&\left(\frac{\partial D_{\omega}}{\partial \bar{\mu}}\right)_T-2 T D_B=0,\\
&\left(\frac{\partial D_{B}}{\partial \bar{\mu}}\right)_T - CT\bar{\mu}  = 0.
\end{align}
The last equation can be integrated trivially
\begin{equation}
D_B(\bar{\mu},T)=\frac{1}{2}CT\bar{\mu}^2+d_B(T).
\end{equation}
Then, from the other equation we get
\begin{equation}
D_\omega(\bar{\mu},T)=\frac{1}{3}CT^2\bar{\mu}^3+2\bar{\mu}Td_B(T)+d_\omega(T).
\end{equation}
From \eqref{dB} we get the equation
\begin{equation}
 T d_B'-d_B =\beta_B.
\end{equation}
This is consistent only if $\left(\frac{\partial \beta_B}{\partial \bar{\mu}}\right)_T=0$, hence $\beta_B$ only depends on the temperature $T$.
Then the equation can be integrated and the general solution is
\begin{equation}
d_B(T)=\gamma_B T +T\int^T \frac{\beta_B(\tau)}{\tau^2}d\tau.
\end{equation}
From \eqref{domega} we get the equation
\begin{equation}
 T d_\omega'-2d_\omega +2\bar{\mu}T\beta_B=\beta_\omega.
\end{equation}
This is a consistent equation only if
\begin{equation}
\beta_\omega=2\bar{\mu}T\beta_B+b_\omega(T).
\end{equation}
The general solution for $d_\omega$ is
\begin{equation}
d_\omega(T)=\gamma_\omega T^2+T^2\int^T \frac{b_\omega(\tau)}{\tau^3}d\tau.
\end{equation}
Finally, we get the following expressions for $\xi_B$ and $\xi_\omega$
\begin{align}
\xi_B & = C\left(\mu - \frac{1}{2}\frac{q \mu^2}{\varepsilon + p}\right) -\frac{q T}{\varepsilon + p}(d_B+\beta_B)\\
\xi_\omega & =  C\left(\mu^2 - \frac{2}{3}\frac{q \mu^3}{\varepsilon + p}\right) +2T \left(d_B-\frac{2\mu q}{\varepsilon+p}d_B\right)-\frac{q T}{\varepsilon+p}(2 d_\omega+ 2T\bar{\mu}\beta_B+b_\omega).
\end{align}

\subsection{Non-relativistic limit in $3+1$ dimensions}\label{entropy1}

The relativistic hydrodynamic equation \eqref{enteq} reduces to the non-relativistic form
\begin{align}
&\partial_t \rho + \partial_i\left(\rho v^i\right) = 0 \\
\label{Ueq} &\partial_t U + \partial_i \left(U v^i\right) -\mu_{NR}\left(\partial_t \rho+\partial_i \left(\rho v^i\right)\right) + p  \partial_i v^i   =  \eta \sigma^2  + \zeta \left( \partial_i v^i\right)^2 + \alpha \left(a^i\right)^2\nonumber \\
& + Ta_i\left(\beta_\omega  \Omega ^i + \beta_B B^i +\frac{2}{T^2}\alpha_2\partial^i T\right) + \partial_i\left(2\alpha_3 a^i + \frac{\sigma}{T}\partial^i T + \xi_\omega \Omega^i + \xi_B B^i\right) - CE^iB_i.
\end{align}
We have defined the vorticity as $\Omega^i=\frac{1}{2}\epsilon^{ijk}\partial_j v_k$.
In this expression we have already extracted the leading order dependence on $c$ of the relativistic coefficients:
\begin{align}
&\eta \sim O\left(c\right), \ \ \zeta \sim O\left(c\right), \ \ \alpha \sim O\left(c^3\right), \ \ \beta_\omega \sim O\left(c^2\right), \ \ \beta_B \sim O\left(c\right), \ \ \alpha_2 \sim O\left(1\right), \ \ \nonumber \\
&\alpha_3 \sim O\left(1\right), \ \ \sigma \sim O\left(\frac{1}{c^3}\right), \ \ \xi_\omega \sim O\left(\frac{1}{c}\right), \ \ \xi_B \sim O\left(\frac{1}{c^2}\right), \ \ C \sim O\left(\frac{1}{c}\right).
\end{align}
In order to find the non-relativistic constraint equations which guarantee positive definite entropy current, we take the non-relativistic limit of (\ref{div3}),
\begin{align}\label{NonRelLim3}
\Delta_{\Pbr} &=O\left(\frac{1}{c}\right)\Omega^i\left( \partial_i D_\omega - \left(2D_\omega + \beta_\omega\right)\frac{ \partial_i p}{\varepsilon + p} - \xi_\omega \partial_i  \left( \frac{\mu}{T} \right) \right) \nonumber \\
&+ O\left(\frac{1}{c^2}\right)\Omega^i E_i \left(\left(2D_\omega + \beta_\omega\right)\frac{q}{\varepsilon + p} - 2D_B + \frac{\xi_\omega}{T}\right) \nonumber \\
&+O\left(1\right) B^i\left(\partial_i D_B - \frac{\partial_i p}{\varepsilon + p}\left(D_B + \beta_B\right) - \xi_B \partial_i\frac{\mu}{T}\right) \nonumber \\
&+O\left(\frac{1}{c}\right)B^i E_i \left(\frac{q \left(D_B + \beta_B\right)}{\varepsilon + p} + \frac{\xi_B}{T} - C\frac{\mu}{T}\right)= 0.
\end{align}
Using the first law of thermodynamics $dU=Tds+\mu_{NR}d\rho$ and 
$U+p=Ts+\mu_{NR}\rho$, Eq.~\eqref{Ueq} determines the divergence of the canonical entropy current in the non-relativistic case. This equation was derived by taking the $O(1/c)$ order of \eqref{enteq} , thus we demand that Eq ~\eqref{NonRelLim3} will be of the same order. This fixes the leading order dependence of $D_\omega$  to be of order  $O\left( c^0 \right)$ and of $D_B$ to be $O\left( \frac{1}{c} \right)$. After factoring out the $c$ dependence and taking the limit, we find the following conditions:
\begin{align}
&\partial_i D_\omega - \frac{\beta_\omega}{\rho}\partial_i p + \frac{\xi_\omega}{T^2}\partial_i T =0 \\ 
&\beta_\omega =0 \\
&\partial_i D_B-\frac{\beta_B}{\rho}\partial_i p + \frac{\xi_B}{T^2}\partial_i T=0\\
& \beta_B - \frac{C}{T}=0
\end{align}

\subsection{Non-relativistic limit in $2+1$ dimensions}\label{entropy2}

The relativistic hydrodynamic equation \eqref{enteq} reduces to the non-relativistic form
\begin{align}
	&\partial_t \rho + \partial_i\left(\rho v^i\right) = 0 \\
	\label{Ueq2}	&\partial_t U + \partial_i \left(U v^i\right) -\mu_{NR}\partial_t \rho - \mu_{NR}\partial_i \left(\rho v^i\right)+ p  \partial_i v^i  \nonumber\\
	&= \eta \sigma^2  + \frac{\zeta}{d} \left( \partial_i v^i\right) + \alpha \left(a^i\right)^2 + \zeta_\omega \partial_k v^k \Omega + \zeta_B B \partial_k v^k+ \frac{1}{T}\left(\tilde\mu_3-T^2\tilde\mu_1 \right) V_i\tilde T^i \nonumber \\
	& + T\tilde\mu_2 \epsilon^{ij} V_i L_j + T\tilde\mu_3 \epsilon^{ij} V_i \partial_j\frac{\mu_{NR}}{T} + \partial_i\left(\left(\frac{\tilde\delta_3}{T}-\tilde\delta_1\right)\tilde T^i + \tilde\delta_2\epsilon^{ij}L_j +\tilde\delta_3\epsilon^{ij}\partial_j\frac{\mu_{NR}}{T}\right).
\end{align}
Where we have defined
\begin{align}
	 \tilde T^i = -\epsilon^{ij}\partial_j T, \ \ L^i = E^i-\epsilon^{ij}v_j B.
\end{align}

In order to find the non-relativistic constraint equations which guarantee positive definite entropy current, we take the non-relativistic limit of (\ref{div2}).  The result is
\begin{align}\label{NonRelLim}
\Delta_{\Pbr} &=O\left(\frac{1}{c}\right)\left(\tilde{\mu}_3-R T^2\tilde{\mu}_1-T\tilde{\delta}_1 -R T^2(\partial_T\tilde{\nu}_5+\tilde{\nu}_1)+(\partial_{\bar{\mu}} \tilde{\nu}_5+\tilde{\nu}_3)+T(\partial_{\bar{\mu}} \tilde{\nu}_1-\partial_T \tilde{\nu}_3)\right) \nonumber \\
& +O\left(\frac{1}{c^3}\right)\left(\tilde{\mu}_2+\frac{1}{T}\tilde{\mu}_3 + \frac{\partial_{\bar{\mu}} \tilde{\nu}_5+\tilde{\nu}_3}{T} + \partial_{\bar{\mu}} \tilde{\nu}_1 - \partial_T \tilde{\nu}_3-T\partial_T \tilde{\nu}_4\right) \nonumber \\
&  + O\left(1\right)\left(\tilde{\delta}_2+\frac{1}{T}\tilde{\delta}_3 + R T (\partial_T \tilde{\nu}_3 -\partial_{\bar{\mu}} \tilde{\nu}_1) - \partial_{\bar{\mu}} \tilde{\nu}_4 + R T^2 \partial_T \tilde{\nu}_4 \right) \nonumber \\
& + O\left(\frac{1}{c}\right) \left(\frac{1}{T}\zeta_B+T\left(\frac{\partial p}{\partial \epsilon}\right)_{q} \partial_T \tilde{\nu}_4 + \frac{1}{T}\left(\frac{\partial p}{\partial q}\right)_{\epsilon}\partial_{\bar{\mu}}\tilde{\nu}_4\right) \nonumber \\
& + O\left(\frac{1}{c^2}\right)\left(\frac{1}{2T}\zeta_\omega + T\left(\frac{\partial p}{\partial \epsilon}\right)_{q} (\partial_T \tilde{\nu}_5 + \tilde{\nu}_1) + \frac{1}{T}\left(\frac{\partial p}{\partial q}\right)_{\epsilon}(\partial_{\bar{\mu}} \tilde{\nu}_5 + \tilde{\nu}_3 )\right) 
\end{align}
Using the first law of thermodynamics $dU=Tds+\mu_{NR}dq$ and $U+p=Ts + \mu_{NR}q$, Eq.~\eqref{Ueq2} determines the divergence of the canonical entropy current in the non-relativistic case. This equation is of order $O(1/c)$, thus we demand that Eq ~\eqref{NonRelLim} will be of the same order. 

Given the expansion of the relativistic chemical potential with $c$ \eqref{cexpq}, in the non-relativistic limit we can replace thermodynamic derivatives to
\begin{align}
&\partial_{\bar\mu}\rightarrow \frac{T}{c}\partial_{\mu_{NR}} \\
&\partial_{T}\rightarrow\partial_{T} + \frac{1}{T}\left(1+\frac{\mu_{NR}}{c^2}\right)\partial_{\mu_{NR}}
\end{align}
Therefore, using also that $R=\frac{1}{c} + O(c^{-3})$,
\begin{align}
&\frac{1}{\rho}\left(U+p\right)\tilde \mu_3^{(0)}+\tilde \mu_3^{(1)} - T^2\tilde \mu_1^{(1)} -T \tilde \delta_1^{(0)} - T^2\partial_T\tilde \nu_5 - T^2 \tilde \nu_1 + \tilde \nu_3 +T^2 \partial_{\mu_{NR}}\tilde \nu_1,  \nonumber \\
&- T \partial_T\tilde \nu_3 - \partial_{\mu_{NR}}\tilde \nu_3 = 0, \\
&\tilde \mu_2 + \frac{1}{T}\tilde \mu_3^{(0)} =0,\\
&\tilde \delta_2 + \frac{1}{T} \tilde \delta_1^{(0)} + T \partial_T \tilde \nu_3 +  \partial_{\mu_{NR}}\tilde \nu_3 - T^2\partial_{\mu_{NR}}\tilde \nu_1 + T^2\partial_T\tilde \nu_4 =0, \\
&\zeta_B =0, \\
&\zeta_\omega =0.
\end{align}
The leading order dependence of the coefficients $\tilde\nu$ with $c$ is,
\begin{align}
\tilde \nu_1 \sim O(c), \ \ \tilde \nu_3 \sim O(1), \ \ \tilde \nu_4 \sim O(1), \ \ \tilde \nu_5 \sim O(c).
\end{align}

\end{document}